\documentclass[12pt]{article}
\usepackage{amsfonts,amssymb}

\newcommand{\Dslash}{{\slash\!\!\!\!D}}
\newcommand{\ootimes}{\mbox{\footnotesize $\otimes$}}

\title{\bf Singular Monopoles\\ via the Nahm Transform}
\author{Sergey A. Cherkis\thanks{E-mail: cherkis@maths.tcd.ie}\\
\it School of Mathematics and Hamilton Mathematics Institute, \\
\it Trinity College, Dublin, Ireland
\rm
\and
Brian Durcan\thanks{E-mail: bdurcan@maths.tcd.ie}\\
\it School of Mathematics,  Trinity College, Dublin, Ireland
}
\date{ }

\begin{document}
\begin{titlepage}

\renewcommand{\thepage}{ }

\maketitle

\begin{abstract}
We present explicit expressions for the fields of a charge one BPS monopole with two Dirac singularities.  These are solutions of the nonlinear Bogomolny equations with the gauge group $U(2)$ or $SO(3).$  We derive these expressions by applying the technique of the Nahm transform.  By exploring various limits we find a number of other solutions.
\end{abstract}

\vspace{-6in}

\parbox{\linewidth}
{\small\hfill \shortstack{TCDMATH 07-23\\ \hfill HMI 07-10}}

\end{titlepage}

\section{Introduction}
Our goal is to find explicit expressions for the fields of the $SO(3)$ and $U(2)$ BPS monopoles \cite{Bogomolny:1975de,Prasad:1975kr} with up to two Dirac singularities. These are pairs $(A, \Phi)$ of the gauge field $A=A_a dx^a$ and the Higgs field $\Phi$ in a three-dimensional flat space satisfying the Bogomolny equation $F=-*D\Phi,$ which in components reads
\begin{equation}
\partial_a A_b-\partial_b A_a-i [A_a, A_b]=-\epsilon_{abc}\big(\partial_c\Phi-i [A_c, \Phi]\big).
\end{equation}
The fields $(A, \Phi)$ are everywhere regular with the exception of a number of specified points $\{\vec{p}_j\}.$ At $\vec{p}_j$ the fields have a Dirac singularity of a given charge $e_j$ imbedded in the gauge group, i.e. the fields approach those of the Goddard-Nuyts-Olive (GNO) monopoles of \cite{Goddard:1976qe}.  To be more specific, there is a gauge in which $\Phi(\vec{x})=e_j \frac{B}{2|\vec{x}-\vec{p}_j|}+O(1),$  where $B$ is an element of the Lie algebra of the gauge group, satisfying ${\rm exp}(2\pi i B)=1.$ We call such a singularity {\em minimal} if  there is no $0<\mu<1$ such that  ${\rm exp}(2\pi i \mu B)=1$ and if $e_j=-1$ or $1.$

We shall focus in this paper on the case of a single nonabelian monopole with one or two minimal singularities.  The exact definition of the nonabelian charge, as given in \cite{Cherkis:1998hi}, corresponds to an intuitive notion of the charge. Namely, for a charge one configuration, whenever the distance between the monopole and the singularities is large we expect the fields  in the vicinity of the monopole to be well approximated by the BPS 't Hooft-Polyakov solution \cite{'tHooft:1974qc, Polyakov:1974ek,Prasad:1975kr} and the fields near each singularity to be approximated by the Dirac-GNO solution.

The main result of this paper is the set of Eqs.~(\ref{Eq:U2fields1}, \ref{Eq:U2fields2}, \ref{SolutionVectors}) expressing the fields of the one $U(2)$ monopole with two oppositely charged singularities.  It is derived using the Nahm transform techniques \cite{Nahm:1982jb, Nahm:1981nb}.

In Section \ref{Sec:Nahm} we outline the Nahm data and the Nahm transform for singular monopoles.  In Section \ref{Sec:U2} we specify the Nahm data  for the charge one case and use the Nahm transform to derive the $U(2)$ solution.
Exploring various limits of this configuration in Section \ref{Sec:Other} we also obtain the following solutions:
\begin{enumerate}
\item a $U(2)$ monopole with two minimal singularities of any charge sign: Eqs.~(\ref{Eq:charges}, \ref{SolutionVectors}), 
\item an $SO(3)$ monopole with two minimal singularities: Eqs.~(\ref{SO(3)}, \ref{SolutionVectors}),
\item a $U(2)$  monopole with one minimal singularity: Eqs.~(\ref{Eq:OneSing}, \ref{Eq:1SingFields}),
\item an $SO(3)$ monopole with one minimal singularity: Eqs.~(\ref{SO(3)}, \ref{Eq:1SingFields}),
\item an $SU(2)$ monopole with one minimal singularity: Eqs.~(\ref{SU(2)}, \ref{Eq:SU(2)fields}).
\end{enumerate}
We conclude with a brief discussion of our results in Section \ref{Sec:Conclusions}.

\section{Nahm Transform}\label{Sec:Nahm}
The Nahm data for singular monopoles is described in detail in \cite{Cherkis:1997aa}.  For a nonabelian charge $k$ $U(2)$ monopole with $k$ positively charged singularities and $k$ negatively charges ones, the Nahm data  consists of two $2 k$-dimensional vectors $Q_{-\lambda}$ and $Q_{\lambda}$  and  four  Hermitian $k\times k$ Nahm matrices  $T_0(s),T_1(s), T_2(s),$ and $ T_3(s)$ defined on the intervals $(-\infty, -\lambda), (-\lambda, \lambda),$ and $(\lambda, +\infty)$ with regular limiting behavior at the boundaries of the corresponding intervals.   We shall often combine the last three of the Nahm matrices into a vector $\vec{T}(s)=(T_1(s), T_2(s), T_3(s)).$   The Nahm equations for the quadruplet $(T_0(s), T_1(s), T_2(s), T_3(s))$ of the rank $k$ Hermitian matrix valued functions read
\begin{equation}
\frac{d}{ds}T_l-i [T_0, T_a]+i \epsilon_{abc} T_b T_c=0,
\end{equation} 
where each of $a,b,$ and $c$ run over the values $1,2$ or $3,$ and the summation over $b$ and $c$ is implied.
The Nahm quadruplet $(T_0, \vec{T})$ transforms under a $U(k)$ valued gauge transformation  $g(s)$ as $(T_0, \vec{T})\mapsto(g^{-1}T_0 g+i g^{-1} dg, g^{-1}\vec{T}g),$ while $Q_{\lambda_\alpha}\mapsto \left(\begin{array}{cc} g^{-1}(\lambda_\alpha)&0\\ 0&g^{-1}(\lambda_\alpha)\end{array}\right) Q_{\lambda_\alpha}.$  We can understand the $Q$'s to be two-spinors in the fundamental representation of the gauge group.  The matching conditions at the points $s=\lambda_\alpha$ (with $\lambda_1=-\lambda$ and $\lambda_2=\lambda$) are the same as in \cite{Hurtubise:1989qy} and can be written as:
\begin{equation}\label{matching}
T_b(\lambda_\alpha+)-T_b(\lambda_\alpha-)=-\frac{1}{2}{\rm tr}_{2\times 2} Q_{\lambda_\alpha} Q^\dagger_{\lambda_\alpha}\sigma_b ,
\end{equation}
where ${\rm tr}_{2\times 2}$ is the trace in the $2$-dimensional spinor space only.

Given the Nahm data satisfying the Nahm equations and the matching conditions (\ref{matching}) for any three-vector $\vec{x},$ we construct the Weyl operator $\Dslash$ defined by
\begin{equation}
{\Dslash} \rho(s)=\left(\begin{array}{c}
\Big(1\ootimes\big(\frac{d}{ds}-i T_0\big)-\vec{\sigma}\ootimes(\vec{T}-\vec{x})\Big)\rho(s)\\
Q_{-\lambda}^\dagger\rho(-\lambda)\\
Q_\lambda^\dagger\rho(\lambda)
\end{array}\right).
\end{equation}
Here $\rho(s)$ is a fundamental two-component spinor-valued section continuous on the real line.
The crucial observation behind the Nahm transform is that the Nahm equations together with the matching conditions are equivalent to
\begin{equation}\label{Laplacian}
{\Dslash}^\dagger{\Dslash}=1\ootimes\bigg(\Big(\frac{d}{ds}-i T_0\Big)^2+T_1^2+T_2^2+T_3^2\bigg),
\end{equation}
and that thus $\Dslash^\dagger\Dslash$ is strictly negative and commutes with the Pauli $\sigma$-matrices.  

The dual Weyl equation is $\Dslash^\dagger \left(\begin{array}{c} \psi\\ \chi\end{array}\right)=0,$ where $\chi=\left(\begin{array}{c} \chi_{-\lambda}\\ \chi_\lambda\end{array}\right),$ reads
\begin{equation}\label{Eq:Weyl}
\bigg(1\ootimes\Big(\frac{d}{ds}-i T_0\Big)+\vec{\sigma}\ootimes(\vec{T}-\vec{x})\bigg)\psi+\delta(s+\lambda) Q_{-\lambda}\chi_{-\lambda}+ \delta(s-\lambda) Q_\lambda \chi_\lambda=0.
\end{equation}
From now on let us denote by $\left(\begin{array}{c} \psi\\ \chi\end{array}\right)$  a matrix with columns being the linearly independent solutions of the dual Weyl equation (\ref{Eq:Weyl}).  There are two square integrable linearly independent solutions of Eq.~(\ref{Eq:Weyl}).  These can be orthonormalized imposing the condition $\int\psi^\dagger\psi ds+\chi^\dagger\chi={\mathbb I}_{2\times 2}.$  The monopole fields are then given by
\begin{eqnarray}\label{NahmTransf1}
\Phi&=&\lambda(\chi_\lambda^\dagger\chi_\lambda- \chi_{-\lambda}^\dagger\chi_{-\lambda})+\int_{-\infty}^{\infty} \psi^\dagger s \psi ds,\\
\label{NahmTransf2}
A_j&=&i \chi^\dagger\frac{\partial}{\partial x^j}\chi+i\int_{-\infty}^{\infty} \psi^\dagger \frac{\partial}{\partial x^j} \psi ds.
\end{eqnarray}
One can note that the rank of the monopole fields is equal to the number of independent solutions of Eq.~(\ref{Eq:Weyl}).

For the purposes of this paper we can restrict ourselves to the Abelian Nahm data only.  In other words, we have $k=1$ and, once $T_0(s)$ is gauge transformed to zero, the Nahm equations imply that the Nahm matrices  $\vec{T}$ are constant on each interval.

\section{One $\mathbf{U(2)}$ Monopole with Two Oppositely Charged Singularities}\label{Sec:U2}
In this section we investigate a monopole solution with a negative charge Dirac singularity at $\vec{p}_1,$ a positive charge Dirac singularity at $\vec{p}_2,$ and the monopole position parameter $\vec{m}.$
We denote the relative positions by $\vec{d}_1=\vec{m}-\vec{p}_1$ and $\vec{d}_2=\vec{m}-\vec{p}_2.$  For an observation point $\vec{x}$ we denote its position with respect to the singularities by $\vec{z}_1=\vec{x}-\vec{p}_1$ and $\vec{z}_2=\vec{x}-\vec{p}_2,$ and its position with respect to the monopole by $\vec{r}=\vec{x}-\vec{m}.$ We also let $z_\alpha=|\vec{z}_\alpha|$ and $d_\alpha=|\vec{d}_\alpha|$ for $\alpha=1,2.$

As mentioned in the previous section, we have $k=1.$ We choose a gauge in which $T_0$ vanishes, then the Nahm equations imply that the Nahm matrices  $\vec{T}$ are $1\times 1$ and that they are constant on each interval. In particular, to produce the monopole specified above, we have
\begin{equation}
\vec{T}=\left\{\begin{array}{ll} 
\vec{p}_1 & {\rm for} \  s<-\lambda,\\
\vec{m} & {\rm for} \  -\lambda<s<\lambda,\\
\vec{p}_2 & {\rm for} \  s>\lambda,
\end{array}\right.
\end{equation}
and the matching conditions (\ref{matching}) become $2 \vec{d}_1=-Q_{-\lambda}^\dagger\vec{\sigma}Q_{-\lambda}$ and $2 \vec{d}_2=Q_{\lambda}^\dagger\vec{\sigma}Q_{\lambda}.$  At this point it is convenient to introduce Weyl spinors $Q_{1\pm}$ and $Q_{2\pm}$ determined by the relations $Q_{\alpha\pm}^\dagger Q_{\alpha\pm}=d_\alpha\pm\vec{\sigma}\cdot\vec{d}_\alpha$ for $\alpha=1,2.$ We have $Q^\dagger_{\alpha\pm} Q_{\alpha\mp}=0.$  
 Also note that these $Q$'s are defined up to a phase factor.  This arbitrariness corresponds to some of the gauge freedom of the final monopole solution.
Now the solutions of the matching conditions are provided by $Q_{-\lambda}=Q_{1-}$ and $Q_\lambda=Q_{2+}.$

Solving the dual Weyl equation is straightforward once we introduce the spinors $\zeta_{\alpha\pm}$ for $\alpha=1,2$ defined by the relations $\zeta_{\alpha\pm}^\dagger\zeta_{\alpha\pm}=z_\alpha\pm\vec{\sigma}\cdot\vec{z}_\alpha.$ The solution to Eq.~(\ref{Eq:Weyl}) then takes the form
\begin{eqnarray}
&\psi(s)=&\left\{\begin{array}{ll} 
e^{\vec{\sigma}\cdot\vec{z}_1(s+\lambda)}\frac{\zeta_{1+} Q_{1+}^\dagger}{Q_{1+}^\dagger\zeta_{1+}} e^{-\vec{\sigma}\cdot\vec{r} \lambda}  N  & {\rm for} \  s<-\lambda\\
e^{\vec{\sigma}\cdot\vec{r}s} N & {\rm for} \  -\lambda<s<\lambda\\
e^{\vec{\sigma}\cdot\vec{z}_2(s-\lambda)}\frac{\zeta_{2-} Q_{2-}^\dagger}{Q_{2-}^\dagger\zeta_{2-}} e^{\vec{\sigma}\cdot\vec{r} \lambda}  N  & {\rm for} \  s>\lambda,
	\end{array}\right. ,\\
&\chi_{-\lambda}=&\frac{\zeta_{1-}^\dagger}{\zeta_{1-}^\dagger Q_{1-}} e^{-\vec{\sigma}\cdot\vec{r}\lambda} N,\\
&\chi_{\lambda}=&-\frac{\zeta_{2+}^\dagger}{\zeta_{2+}^\dagger Q_{2+}} e^{\vec{\sigma}\cdot\vec{r}\lambda} N.
\end{eqnarray}
The matrix $N$ is independent of the variable $s.$ The  ortonormalization condition implies that
\begin{displaymath}
N=\frac{\sqrt{r}}{2}\frac{\Big(\sqrt{z_1+d_1+\sqrt{{\cal D}_1}}+\sqrt{z_1+d_1-\sqrt{{\cal D}_1}}\frac{\vec{\sigma}\cdot\vec{r}}{r}\Big)\Big(\sqrt{z_2+d_2+\sqrt{{\cal D}_2}}-\sqrt{z_2+d_2-\sqrt{{\cal D}_2}}\frac{\vec{\sigma}\cdot\vec{r}}{r}\Big)}
{\sqrt{{\Big( (z_1+d_1)(z_2+d_2)+r^2\Big)\sinh(2\lambda r)+r(z_1+d_1+z_2+d_2)\cosh(2\lambda r)}}},
\end{displaymath}
where for brevity we introduced ${\cal D}_1=(z_1+d_1)^2-r^2$ and ${\cal D}_2=(z_2+d_2)^2-r^2.$  The normalization condition restricts $N$ up to an $s$-independent $U(2)$ transformation acting on the right.  Multiplying the normalization matrix $N$ above on the right by an element of $U(2)$ would  amount to a gauge transformation of the resulting monopole fields.

Introducing the following simple functions
\begin{eqnarray}
{\frak  L}&=& ((z_1+d_1) (z_2+d_2)+r^2)\sinh( 2\lambda r)+r(z_1+z_2+d_1+d_2)\cosh(2\lambda r),\\
{\frak K}&=& ((z_1+d_1) (z_2+d_2)+r^2)\cosh( 2\lambda r)+r(z_1+z_2+d_1+d_2)\sinh(2\lambda r),
\end{eqnarray}
we use Eqs.~(\ref{NahmTransf1},\ref{NahmTransf2}) to obtain the following monopole solution:
\begin{eqnarray}\label{Eq:U2fields1}
\Phi&=&\frac{1}{4z_2}-\frac{1}{4z_1}+\vec{\sigma}\cdot\vec{\phi},\\
\label{Eq:U2fields2}
A&=&\frac{(\vec{r}\times\vec{z}_1)\cdot d\vec{x}}{2 z_1{\cal D}_1}
-\frac{(\vec{r}\times\vec{z}_2)\cdot d\vec{x}}{2 z_2{\cal D}_2}+\vec{\sigma}\cdot\vec{A},
\end{eqnarray}
where $\vec{\phi}$ and $\vec{A}$ are
\begin{eqnarray}
\label{SolutionVectors}
\vec{\phi} &=&\frac{\vec{r}}{r}\left(\bigg(\lambda+\frac{1}{4z_1}+\frac{1}{4z_2}\bigg)\frac{{\frak K}}{{\frak L}}-\frac{1}{2r}\right)-
\frac{r\sqrt{{\cal D}_1{\cal D}_2}}{{2\frak L}}\left(\frac{\vec{d}_{1\perp}}{z_1 {\cal D}_1}+\frac{\vec{d}_{2\perp}}{z_2 {\cal D}_2}\right),
\\
\vec{A}&=& \frac{\vec{r}\times d\vec{x}}{r}\Big(\big(\lambda+\frac{ z_1+d_1}{2{\cal D}_1}+\frac{ z_2+d_2}{2{\cal D}_2}\big)\frac{\sqrt{{\cal D}_1{\cal D}_2}}{{\frak L} }-\frac{1}{2r}\Big)-\frac{(\vec{r}\times\vec{z}_1)\cdot d\vec{x}}{2 z_1{\cal D}_1}\frac{\vec{r}}{r}\frac{{\frak K}}{{\frak L} }
-\frac{(\vec{r}\times\vec{z}_2)\cdot d\vec{x}}{2 z_2{\cal D}_2}\frac{\vec{r}}{r}\frac{{\frak K}}{{\frak L} }\nonumber\\
&&-\frac{r\sqrt{{\cal D}_1{\cal D}_2}}{{\frak L} }\Big(
\frac{\vec{\sigma}\cdot(\vec{z}_1\times d\vec{x})_\perp}{2 z_1{\cal D}_1}
+\frac{\vec{\sigma}\cdot(\vec{z}_2\times d\vec{x})_\perp}{2 z_2{\cal D}_2}\Big).\nonumber
\end{eqnarray}
Here $(\vec{v}\times d\vec{x})_\perp$ is the projection of the vector $\vec{v}\times d\vec{x}$ on the plain orthogonal to $\vec{r},$ i.e.
$
(\vec{v}\times d\vec{x})_\perp=\vec{v}\times d\vec{x}-((\vec{v}\times d\vec{x})\cdot\vec{r})\frac{\vec{r}}{r^2}.
$

\section{Other Solutions}\label{Sec:Other}
\subsection{$\mathbf{U(2)}$ Monopole with Two Singularities of Charges $\mathbf{\pm1}$}
This solution is obtained by adjusting the sign of the Abelian (trace) part of the singularities.
\begin{eqnarray}\label{Eq:charges}
\Phi&=&\frac{e_1}{4z_1}+\frac{e_2}{4z_2}+\vec{\sigma}\cdot\vec{\phi},\\
A&=&-\frac{e_1}{2 z_1{\cal D}_1}(\vec{r}\times\vec{z}_1)\cdot d\vec{x}
-\frac{e_2}{2 z_2{\cal D}_2}(\vec{r}\times\vec{z}_2)\cdot d\vec{x}
+\vec{\sigma}\cdot\vec{A},\nonumber
\end{eqnarray}
where the nonabelian parts of the fields $\vec{\phi}$ and $\vec{A}$ are still given by Eq.~({\ref{SolutionVectors}).
We note that deriving this solution via the Nahm transform is more challenging, since in this case the required Nahm data is nonabelian.

\subsection{$\mathbf{SO(3)}$ Monopole Solution with Two Singularities}
Dividing out the center of the $U(2)$ one is led to the $SO(3)$ solution
\begin{equation}\label{SO(3)}
\Phi_{ij}=-2i\epsilon_{ijk}\phi^k,  A_{ij}=-2i \epsilon_{ijk} A^k,
\end{equation}
with the vectors $\vec{\phi}=(\phi^1,\phi^2,\phi^3)$ and $\vec{A}=(A^1,A^2,A^3)$ still given by Eq.~({\ref{SolutionVectors}).

Now we proceed to study the two natural limits of these solutions.  

\subsection{$\mathbf{U(2)}$ and $\mathbf{SO(3)}$ Monopoles with One Minimal Singularity}
A solution for a $U(2)$ monopole with only one minimal Dirac singularity is obtained by considering the limit with $\vec{p}_2\rightarrow\infty,$ leading to  
\begin{equation}\label{Eq:OneSing}
\Phi=\frac{e}{4z}+\vec{\sigma}\cdot\vec{\phi},\quad 
A=-\frac{e}{2 z{\cal D}}(\vec{r}\times\vec{z})\cdot d\vec{x}+\vec{\sigma}\cdot\vec{A},
\end{equation}
with
\begin{eqnarray}\label{Eq:1SingFields}
\vec{\phi}&=&\left(\bigg(\lambda+\frac{1}{4 z}\bigg)\frac{K}{L}-\frac{1}{2 r}\right)\frac{\vec{r}}{r}-\frac{r}{2 z L \sqrt{{\cal D}}}\left(\vec{d}-\frac{\vec{r}\cdot\vec{d}}{r^2}\vec{r}\right),\\
\vec{A}&=&\left(\bigg(\lambda+\frac{z+d}{2{\cal D}} \bigg)\frac{\sqrt{{\cal D}}}{L}-\frac{1}{2r}\right)\frac{\vec{r}\times d\vec{x}}{r}
-\frac{r}{2 L \sqrt{{\cal D}}}\left(\frac{\vec{z}\times d\vec{x}}{z}+ \bigg(\frac{K}{\sqrt{\cal D}}-1 \bigg) \frac{(\vec{r}\cdot(\vec{z}\times d\vec{x}))}{r z}\frac{\vec{r}}{r}\right).\nonumber
\end{eqnarray}
Here we make use of the function ${\cal D}=(z+d)^2-r^2,$ as well as
\begin{eqnarray}
K&=&(z+d)\cosh(2\lambda r)+r\sinh(2\lambda r),\\
L&=&(z+d)\sinh(2\lambda r)+r\cosh(2\lambda r).
\end{eqnarray}
These expressions reproduce the first  solution of \cite{Cherkis:2007jm}, which was obtained in \cite{thesis} directly via the Nahm transform using a different set of corresponding Nahm data.

An $SO(3)$ BPS monopole with one singularity is given by Eqs.~(\ref{SO(3)}) with $\vec{A}$ and $\vec{\phi}$ as given above in Eqs.~(\ref{Eq:1SingFields}).

\subsection{Coincident Singularities}
The limit of coinciding singularities $\vec{p}_2\rightarrow\vec{p}_1$ produces a pure $SU(2)$ solution:
\begin{equation}\label{SU(2)}
\Phi=\vec{\sigma}\cdot\vec{\phi},\quad A=\vec{\sigma}\cdot\vec{A},
\end{equation} 
with
\begin{eqnarray}\label{Eq:SU(2)fields}
\vec{\phi}&=&\left(\bigg(\lambda+\frac{1}{2z} \bigg)\frac{\cal K}{{\cal L}}-\frac{1}{2r}\right)\frac{\vec{r}}{r}-\frac{r}{z{\cal L}}\left(\vec{d}-\frac{\vec{r}\cdot\vec{d}}{r^2}\vec{r}\right),\\
\nonumber
\vec{A} &=&\left(\bigg(\lambda+\frac{z+d}{{\cal D}} \bigg)\frac{{\cal D}}{{\cal L} }-\frac{1}{2r}\right)\frac{\vec{r}\times d\vec{x}}{r}
-\frac{r}{{\cal L}}\left(\frac{\vec{z}\times d\vec{x}}{z}+ \bigg(\frac{{\cal K}}{{\cal D}}-1 \bigg)\frac{(\vec{r}\cdot(\vec{z}\times d\vec{x}))}{rz}\frac{\vec{r}}{r}\right),
\end{eqnarray}
where
\begin{eqnarray}
\nonumber
{\cal K}&=&((z+d)^2+r^2)\cosh(2\lambda r) + 2r(z+d)\sinh(2\lambda r),\\
\nonumber
{\cal L}&=&((z+d)^2+r^2)\sinh(2\lambda r) + 2r(z+d)\cosh(2\lambda r).
\end{eqnarray}
This is the second solution analyzed in \cite{Cherkis:2007jm}.

\section{Conclusions}\label{Sec:Conclusions}
The solutions presented here can be used in a number of physical contexts, some of which we now discuss.

The Dirac-GNO singularities have the interpretation as 't~Hooft operators \cite{'t Hooft:1977hy}. The 't~Hooft operators are the disorder parameter fields of the QCD and the behavior of their Green's functions has direct bearing on the question of confinement.  The leading classical terms of these Green's functions is delivered by the classical tension of the Dirac string.  The monopole configuration described here provides one of the nonperturbative corrections contributing, for example, at finite temperature as well as at the finite volume.
Lifting the two singularity $SO(3)$ solution of Eqs.~(\ref{SO(3)},\ref{SolutionVectors}) to a configuration in the $SU(2)$ gauge theory results in a configuration containing a physical Dirac string connecting the two singular points.  From the four-dimensional point of view it describes a nonabelian monopole in  $SU(2)$ in the presence of two world-lines of the Dirac singularities. The Dirac monopoles  are the genuine 't~Hooft line operator in the original sense of \cite{'t Hooft:1977hy}, and the world-sheet of the Dirac string connecting them signifies a physical interface between the two vacua related by the $\mathbb{Z}_2$ symmetry of the center of the gauge group.  

The significance of 't~Hooft operators for the Montonen-Olive duality \cite{Montonen:1977sn} is elucidated in \cite{Kapustin:2005py} where they emerge as S-dual partners of the Wilson-Polyakov operators.  't~Hooft operators also play significant role as Hecke operators in the supersymmetric gauge theory realization of the Langlands duality \cite{Kapustin:2006pk}.

The screening effects of the singularities by the monopole are also of interest. It is akin to the instanton bubbling effect.  In \cite{Cherkis:2007jm} we demonstrate that a $U(2)$ monopole cannot completely screen a single minimal singularity, however, we find that it can screen completely two coinciding singularities of opposite charge.  Solutions presented in this paper provide a continuous family interpolating these two extreme cases.  It would be interesting to explore the `partial screening' of a pair of closely positioned singularities.  One can use such screening to define 't~Hooft-like operators localized in space-time.  Namely, if a pair of coinciding 't~Hooft operators can be screened by a monopole, one can also create such a pair and a monopole by moving the monopole away at some time $t_i$ and then rejoin the monopole and the singularity at a later time $t_f.$  In between $t_i$ and $t_f$ one can also separate the two singularities from each other at $t_1$ and rejoin them again at $t_2>t_1>t_i,$ thus creating a conventional 't~Hooft loop between the times $t_1$ and $t_2$ within the monopole-singulatiry loop.

Our last comment is regarding the moduli spaces of these monopoles.  All the monopoles discussed here have a four dimensional moduli space.  The four coordinates on this space are given by the position of the monopole in the three-dimensional space and its phase in the gauge group.  As discussed in \cite{Cherkis:1998hi} the moduli space of one $U(2)$ monopole with two minimal singularities is a two-centered Taub-NUT space. (Same holds for the $SO(3)$ monopole with two singularities.) Thus, following the logic of \cite{Manton:1985hs,Gibbons:1995yw}, the low velocity scattering of this monopole off a pair of singularities is given by a geodesic motion on the two-centered Taub-NUT.

The moduli space of one $U(2)$ or $SO(3)$ monopole with a single singularity is the Taub-NUT space.  The moduli space of one $SU(2)$ monopole with one singularity is a degenerate Taub-NUT space, which can be viewed as Taub-NUT$/\mathbb{Z}_2.$  

\section*{Acknowledgments}
BD is supported by the Irish Research Council for Science Engineering and Technology: funded by the National Development Plan. The work of SCh is supported by the Science Foundation Ireland Grant No. 06/RFP/MAT050 and by the European Commision FP6 program MRTN-CT-2004-005104.

\end{document}